\documentstyle[aps,prb]{revtex}
\begin{document}

\title{Upper critical field in layered superconductors}
\author{V.P.Mineev$^{1,2}$}
\address{$^1$ Yukawa Institute for Theoretical Physics, Kyoto 
University,
Kyoto 606-8502, Japan}
\address{$^2$ Commissariat a l'Energie Atomique, Departement de 
Recherche Fondamentale sur la Matiere Condensee, SPSMS, 38054, 
Grenoble, France }
\date{Submitted 23 February 2000}
\maketitle

\begin{abstract}
The theoretical statements about a restoration of a superconductivity 
at magnetic fields higher than the quasiclassical upper critical 
field and a reentrance of superconductivity at temperatures 
$T_c(H)\approx T_c(0)$
in the superconductors with open Fermi surfaces are reinvestigated 
taking into
account a scattering of quasiparticles on the impurities.

The system of integral equations for determination of the upper 
critical field parallel to the conducting planes in a layered 
conventional and unconventional superconductors with impurities are 
derived. The $H_{c2}(T)$ values for the
"clean" case in the Ginzburg-Landau regime and at any temperature in 
the "dirty"
case are found analytically.
The upper limit of the superconductor purity when the upper critical 
field definately has a finite value is established.

\end{abstract}

\newpage

\section{Introduction}

The theory of upper critical field in highly anisotropic 
quasi-two-dimensional superconductors for the field orientation 
parallel to conducting layers has been developed  by A.Lebed' and 
K.Yamaji \cite{1}. It was shown that like in quasi-one-dimentional 
case \cite{2} at low enough
temperatures (see below) the upper critical field starts to diverge 
such that a superconductivity is conserved in arbitrary large 
magnetic fields. Moreover in high enough magnetic fields (see below) 
the critical temperature of a superconducting phase transition has a 
tendency to restore of its zero field value. These statements are 
literally true for the superconductors with triplet pairing. 
For a singlet pairing the existance of a superconductivity in high 
fields is restricted by the paramagnetic limit $H_p$. So, the low 
temperature stability of
a superconducting state under magnetic field can serve as the 
indication on the superconductivity with triplet pairing.

 The tendency for low temperature divergence of the upper critical 
field $H_{c2}>H_p$  in quasi-two-dimentional 
$\kappa$-type ET organic superconductors has been reported recently by T.Ishiguro 
\cite{3}. Similar behavior has been observed earlier
in quasi-one-dimensional organic superconductor $(TMTSF)_2PF_6$ by 
the group
of P.Chaikin \cite{4}. So, the observations being in correspondence 
with
theoretical predictions \cite{1,2}
say in favor of triplet type of superconductivity in both type of 
these materials. 

Another popular layered superconductor is $Sr_2RuO_4$.
It demonstrates the properties compatible with triplet 
superconductivity \cite{5}.
Unlike isostructural high-$T_c$ cuprates $La_{2-x}Ba_xCuO_4$ the 
normal state of layered perovskite oxide $Sr_2RuO_4$ conforms with 
the predictions of Fermi-liquid theory  \cite{6}. Unconventional 
nature of the superconducting state in this material manifests itself 
through
the strong supression of $T_c$ by nonmagnetic impurities \cite{7,8} 
as well as by the exibition  of a sharp decrease without the 
coherence peak of $^{101}Ru$ nucleare spin-relaxation rate $1/T_1$ 
followed by $T^3$ behavior down to $0.15 K$ \cite{9}.  The spin part 
of $^{17}O$ Knight shift  for the field $H=0.65 T$ parallel to ${\it 
ab}$ plane does not change down to $15 mK$, much below 
$T_c(H)\approx 1.2 K$ \cite{10}. Since the experiment has been 
performed on $Sr_2RuO_4$ in a clean limit this constitues the 
evidence for the triplet spin pairing  with spin of paires lying in 
${\it ab}$ plane. On the other hand the 
measurements of the basal plane upper critical field down to $0.2 K$ 
shows no tendency to divergency of $H_{c2}(T)$ \cite{11} ( see also 
results of measurements on less perfect crystals \cite{12}). 
Moreover, the search for reentrance of superconductivity under 
magnetic fields up to $33 T$
and temperatures down to {50 mK} has given the negative results 
\cite{13}.
The basal plane upper critical field saturates at low temperatures at 
$1.5 T$
which is well below the paramagnetic limit $H_p\approx 2.8 T$ and 
roughly corresponds to quasiclassical upper critical field value
\begin{equation}
H_{c2}=\frac{\Phi_0}{2\pi\xi_{ab}\xi_{c}},
\label{e1}
\end{equation}
where $\Phi_0$ is the flux quantum and $\xi_{ab}$, $\xi_{c}$ are the 
coherence 
lengths in basal plane and along the $c$ axis correspondingly.

These observations being certainly in contradiction with theory 
\cite{1} stimulate us to reinvestigate the problem of upper critical 
field in quasi-two-dimensional superconductors taking into account 
scattering of
quasiparticles on the impurities. The main goal is to find a limits 
for crystal purity when one can hope to see the low temperature upper 
critical field divergency.
In a few words the result can be described as follows. In a pure 
crystal the $H_{c2}$ divergency
starts to be developed when the thermal coherence length $\xi (T)= 
v_F/2\pi T$ begins to be
larger than the "cyclotron radius" of the quasiparticles orbits 
$R_c(H)=v_F/\omega_c$. Here $v_F$ is the basal plane Fermi velocity,
$\omega_c=eHv_Fd/c$ is the "cyclotron frequency", $d$ is the distance 
between
conducting layers. 
 The impurities does not prevent the upper critical field divergency 
if at the temperature determined by equality
\begin{equation}
\xi(T)=R_{c}(H_{c2})
\label{e2}
\end{equation}
the quasiparticles mean free path $l$ is still larger than $\xi(T)$.  
Otherwise
the upper critical field is saturated at temperatures below  
\begin{equation}
T_l=\frac{v_F}{2\pi l}
\label{e3}
\end{equation}
and its value is roughly described by formula 
(\ref{e1}).

The paper has following structure. The general formalism for the 
upper critical field problem in layered conventional and 
unconventional superconductors is introduced in the next Section. By 
the way of the derivation the several simplifications have been used: 
(i) The equations are written for axially symmetric crystal where the 
only one-dimentional, even or odd in respect to the reflections
in (${\bf H}$, crystal axis) plane superconducting states are 
realized;
(ii) Among them the only even superconducting states are considered;
(iii) The equations are derived with but solved in neglect Pauli 
paramagnetic
interaction;
(iv)The equations are solved for only unconventional superconducting 
states with zero anomalous Green function self-energy. The analytical 
solution of the equations
accompanied by the discussion of limits of crystal purity sufficient 
for the 
upper critical field saturation at low temperature is containned in 
the third Section.

\section{The order parameter equations}

The electron spectrum of a layered crytals  obeys basal plain 
anizotropy,  $z$-dependent corrugation of the Fermi surface and 
several bands in general case. It seems however unimportant for our 
purposes to take into account all these complifications. 
So, we shall consider a metal with
electron spectrum
\begin{eqnarray}
&\epsilon&({\bf p})=\frac{1}{2m}(p_x^2 +p_y^2)-2t\cos p_zd, \nonumber 
\\
&t&\ll\epsilon_F=\frac{mv_F^2}{2}, \qquad 
\frac{-\pi}{d}<p_z<\frac{\pi}{d}
\label{e4}
\end{eqnarray}
in the magnetic field ${\bf H}=(0,H,0)$, ${\bf A}=(0,0,-Hx)$ parallel
to the conducting layers with distance $d$ between them. 
In absence of impurities the normal state electron Green function 
$G_{\omega_n,\sigma}(p_y,p_z,x,x')$
is obtained as the result of solution of the equation
\begin{equation}
\left [i\omega-\frac{1}{2m}\left (-\frac{d^2}{dx^2}+p_y^2 \right)+
2t\cos \left ( p_zd-\frac {\omega_c x}{v_F}\right) +\sigma\mu_e H+\mu 
\right ]
G_{\omega,\sigma}(p_y,p_z,x,x')=\delta(x-x'),
\label{e5}
\end{equation}
where $\omega_n=\pi T(2n+1)$ is the Matsubara frequency, 
$\omega_c=ev_FdH/c$,
$v_F$ is the Fermi velocity, 
$\hbar=1$, $\mu$ is the chemical potential, $\mu_e$ is a magnetic 
moment of an electron in a
crystal. To use the diagonal shape of the Green function matrix we 
have chosen
the $\hat y$ direction as the spin quantization axis, such that 
$\sigma=\pm 1$.
 
In a layered crystal with a singlet Cooper pairing the 
superconducting states obey the following order parameters
\begin{equation}
\hat \Delta^{s}•=i\Delta^{s}•(p_x,p_y,p_z, x)\hat \sigma_y,
\label{e6}
\end{equation}
Here, $\hat \sigma_y$
is the Pauli matrix. 
For a triplet superconductivity we shall limit ourselves by the 
consideration of the so called equal spin pairing 
states with spin lying in the plane of the conducting layers.
In neglect of small effects of spontaneous magnetism \cite{14}
a vector wave function of such the states is
\begin{equation}
{\bf d}=\hat z \Delta^{t}•(p_x,p_y, p_z, x).
\label{e7}
\end{equation}
As we have put the spin quantization axis along $\hat y$ direction we 
will use the corresponding basis of Pauli matrices 
$\overrightarrow\sigma=(\hat\sigma_y,\hat\sigma_z,\hat\sigma_x)$.
So the order parameter for triplet pairing state in our case is
\begin{equation}
\hat \Delta^{t}•=i({\bf d}\overrightarrow\sigma)\sigma_y=
-\Delta^{t}•(p_x, p_y, p_z, x)\hat \sigma_z.
\label{e8}
\end{equation}  

The order parameter function is represented \cite{15} as a linear 
combination of the basis
functions $\psi_i(\phi, p_z)$ of one of the irreducible 
representations of crystal symmetry group 
\begin{equation}
\Delta^{s,t}•(p_x,p_y, p_z, x)=\psi_i(\phi, p_z)\Delta^{s,t}•_i(x).
\label{e9}
\end{equation}
Here, $\phi$ is the angle between basal plane vector of momentum 
${\bf p}_{\parallel}$
and magnetic field ${\bf H}\parallel \hat y$.  There are only one and 
two dimensional representations ($i=1,...d; d=1,2$) in the crystals 
with axially symmetric spectrum (\ref{e4}).

The  
crystal with
axial symmetry under magnetic field lying in the basal plane
obeys the symmetry in respect of reflections in plane where 
the vectors of magnetic field and crystal axis lie,
that is $(y,z)$ plane in our case \footnote{For the uniaxial crystals with 
hexagonal or tetragonal
symmetry this property takes place only for particular directions of 
magnetic field
in the basal plane.}. The two components of vector basis functions 
$(\psi_1(\phi, p_z), \psi_2(\phi, p_z))$ can be always chosen such 
that each of them will have definite (even or odd) and at the same 
time mutually opposite parity in respect to reflections in $(y,z)$ 
plane.
As the consequence the set of the order parameter equations for two 
component superconductivity splits on two independent equation for 
each component of the order parameter.
One of them corresponds to the higher value of the upper critical 
field. Thus we always deal with one-component superconductivity with 
definite parity. The simplest examples
of even functios $\psi(\phi, p_z)$ are:
$1,~\sin ^2\phi-\cos ^2\phi~~$ for singlet pairing 
and $ ~\sin(p_zd),~\cos \phi$ for triplet pairing. As an examples of 
odd states one can pointed out on  $ ~\sin(p_zd)\sin \phi$ for 
singlet pairing and $\sin \phi$ for the triplet pairing.

The upper critical field is found from the equation on the order 
parameter
which have the different shape for even and odd superconducting 
states. For determiness we shall consider just the even order 
parameter states. 
In this case the order parameter equation for a clean superconductor 
with singlet pairing is
\begin{eqnarray}
&\Delta^{s}•(\phi, p_z,x)&=g \int dx' 
\int \frac{dp'_y}{2\pi} 
\int \limits_{-\pi/d}^{\pi/d}\frac{dp'_z}{2\pi}
\psi(\phi, p_z)\psi^*(\phi', p'_z)\nonumber \\
&\times&T\sum_n \frac{1}{2} \sum_{\sigma=\pm 1} 
G_{\omega_n,\sigma}(p'_y,p'_z,x,x')
G_{-\omega_n,-\sigma}(-p_y',-p'_z,x,x')
\Delta^{s}•(\phi', p'_z,x')
\label{e10}
\end{eqnarray}
and for the triplet pairing
\begin{eqnarray}
&\Delta^{t}•(\phi, p_z,x)&=g\int dx' 
\int \frac{dp'_y}{2\pi} 
\int \limits_{-\pi/d}^{\pi/d}\frac{dp'_z}{2\pi}
\psi(\phi, p_z)\psi^*(\phi', p'_z)\nonumber \\
&\times& T\sum_n\frac{1}{2} \sum_{\sigma=\pm 1}
G_{\omega_n,\sigma}(p'_y,p'_z,x,x')
G_{-\omega_n,\sigma}(-p'_y,-p'_z,x,x')
\Delta^{t}•(\phi', p'_z,x')
\label{e11}
\end{eqnarray}

In presence of the impurities the order parameter of a 
superconducting 
singlet pairing state  \linebreak 
$\hat \Delta^{s}•=i\Delta^{s}•(\phi, p_z,x)\hat \sigma_y$
acquires a self energy part 
\begin{equation}
\hat \Sigma^{s}•(\omega_{n}•,x)
=i\Delta^s_{\omega_n}(x)\hat \sigma_y+\Delta^t_{\omega_n}(x)\hat \sigma_x
\label{e12}
\end{equation} 
consisting of singlet and triplet components \footnote {Compare with 
the paper \cite{16},
where a similar theory in frame of quasiclasical approach has been 
developed.}

For a triplet pairing states with an order parameter 
$\hat \Delta^{t}•=-\Delta^{t}•(\phi, p_z, x)\hat \sigma_z$
the  corresponding self energy part 
\begin{equation}
\hat \Sigma^{t}•(\omega_{n}•,x)=
-\Delta^t_{\omega_n}(x)\hat \sigma_z+i\tilde\Delta^t_{\omega_n}(x)\hat \sigma_0
\label{e13}
\end{equation} 
consists of two different triplet components. Here $\hat \sigma_0$ is 
the two dimensional unit matrix. The singlet component of self energy
part is absent for chosen basal plane orientation of magnetic field 
and equal spin
triplet pairing with spin directions parallel to conducting layers. 

The self-consistency equations for the 
order parameters and self energy parts have the form (see \cite {15})
\begin{eqnarray}
&\Delta_{\alpha \beta}•^{s,t}•&= \int dx'
\int\frac{dp'_y}{2\pi} 
\int \limits_{-\pi/d}^{\pi/d}\frac{dp'_z}{2\pi}
V_{\beta \alpha,\lambda \mu}•^{s,t}•\nonumber \\
&\times& T\sum_n
G_{\tilde \omega_n}^{\lambda \gamma}•(p'_y,p'_z,x,x')
G_{-\tilde \omega_n}^{\mu \delta}•(-p'_y,-p'_z,x,x')
[\Delta^{s,t}•_{\gamma \delta}•(\phi', p'_z,x')+
\Sigma^{s,t}•_{\gamma \delta}(\tilde\omega_n,x')],\nonumber \\
&\Sigma_{\gamma \delta}^{s,t}•&(\tilde\omega_n,x)= nu^{2}•\int dx'
\int\frac{dp'_y}{2\pi} 
\int \limits_{-\pi/d}^{\pi/d}\frac{dp'_z}{2\pi} \nonumber \\
&\times& G_{\tilde \omega_n}^{\gamma \alpha }•(p'_y,p'_z,x,x')
G_{-\tilde \omega_n}^{\beta \delta}•(-p'_y,-p'_z,x,x')
[\Delta^{s,t}•_{\alpha \beta}•(\phi', p'_z,x')+
\Sigma^{s,t}•_{\alpha \beta}(\tilde\omega_n,x')],
\label{e14}
\end{eqnarray}
where $V_{\beta \alpha,\lambda \mu}•^{s,t}•=g~g^{s,t}•_{\beta \alpha}•
g^{s,t~+}•_{\lambda \mu}•\psi(\phi, p_z)\psi^*(\phi', p'_z)/2$,
$~\hat g^{s}•=i\hat \sigma_{y }•$, $~\hat g^{t}•=-\hat 
\sigma_{z }•$; \\ 
$G_{\tilde \omega_n}^{\lambda \gamma}•=G_{\tilde \omega_n,1}
(\sigma_{0}•^{\lambda \gamma}•+\sigma_{z}•^{\lambda \gamma}•)/2+
G_{\tilde \omega_n,-1}
(\sigma_{0}•^{\lambda \gamma}•-\sigma_{z}•^{\lambda \gamma}•)/2$. \\
Here 
$\tilde \omega_n=\omega_n+\Gamma{\it sign}\omega_n$, 
$\Gamma=mnu^2/2d$ and $u$, $n$ are impurity potential amplitude and 
concentration.
We will use also a quasiparticle life time $\tau$ and a mean free 
path $l$ introduced
by means $\Gamma=1/2\tau=v_F/2l$.
  
The equations (\ref{e14}) are obtained in frame of 
 procedure of averaging over an impurity positions.  As it was
 shown in the paper \cite {17} one can use 
 the field independent value of $\Gamma$ so long
 \begin{equation}
\frac{v_F}{\omega_c}>\frac{l}{\sqrt{k_{F}•}l}.
\label{e15}
\end{equation}

The system of "the scalar" self-consistency equations for the singlet pairing 
states following of equations (\ref{e14}) is
\begin{eqnarray}
&\Delta^{s}•&(\phi, p_z,x)=g \int dx'
\int\frac{dp'_y}{2\pi} 
\int \limits_{-\pi/d}^{\pi/d}\frac{dp'_z}{2\pi}
\psi(\phi, p_z)\psi^*(\phi', p'_z)\nonumber \\
&\times& T\sum_n\frac{1}{2} \sum_{\sigma=\pm 1}
G_{\tilde \omega_n,\sigma}(p'_y,p'_z,x,x')
G_{-\tilde \omega_n,-\sigma}(-p'_y,-p'_z,x,x')
[\Delta^{s}•(\phi', p'_z,x')+\Delta^s_{\tilde\omega_n}(x')+
\sigma \Delta^t_{\tilde\omega_n}(x')],
\label{e16}
\end{eqnarray}
\begin{eqnarray}
&\Delta&^s_{\tilde\omega_n}(x)=nu^2\int dx' 
\int\frac{dp_y}{2\pi} 
\int \limits_{-\pi/d}^{\pi/d}\frac{dp_z}{2\pi} 
\frac{1}{2} \sum_{\sigma=\pm 1}
G_{\tilde \omega_n,\sigma}(p_y,p_z,x,x')
G_{-\tilde \omega_n,-\sigma}(-p_y,-p_z,x,x')\nonumber \\
&\times&[\Delta^{s}•(\phi,\hat p_z,x')+\Delta^s_{\tilde\omega_n}(x')+
\sigma 
\Delta^t_{\tilde\omega_n}(x')],
\label{e17}
\end{eqnarray}
\begin{eqnarray}
&\Delta&^t_{\tilde\omega_n}(x)=nu^2\int dx' 
\int \frac{dp_y}{2\pi} 
\int \limits_{-\pi/d}^{\pi/d}\frac{dp_z}{2\pi} 
\frac{1}{2} \sum_{\sigma=\pm 1}
G_{\tilde \omega_n,\sigma}(p_y,p_z,x,x')
G_{-\tilde \omega_n,-\sigma}(-p_y,-p_z,x,x')\nonumber \\
&\times&[\sigma (\Delta^{s}•(\phi,\hat 
p_z,x')+\Delta^s_{\tilde\omega_n}(x'))+
\Delta^t_{\tilde\omega_n}(x')].
\label{e18}
\end{eqnarray}
For the triplet pairing case the corresponding equations are
\begin{eqnarray}
&\Delta^{t}•&(\phi, p_z,x)=g \int dx' 
\int \frac{dp'_y}{2\pi} 
\int \limits_{-\pi/d}^{\pi/d}\frac{dp'_z}{2\pi}
\psi(\phi, p_z)\psi^*(\phi', p'_z)\nonumber \\
&\times& T\sum_n\frac{1}{2} \sum_{\sigma=\pm 1}
G_{\tilde \omega_n,\sigma}(p'_y,p'_z,x,x')
G_{-\tilde \omega_n,\sigma}(-p'_y,-p'_z,x,x')
[\Delta^{t}•(\phi', p'_z,x')+\Delta^t_{\tilde\omega_n}(x')
-i\sigma 
\tilde\Delta^t_{\tilde\omega_n}(x')],
\label{e19}
\end{eqnarray}
\begin{eqnarray}
&\Delta&^t_{\tilde\omega_n}(x)=nu^2 \int dx' 
\int \frac{dp_y}{2\pi} 
\int \limits_{-\pi/d}^{\pi/d}\frac{dp_z}{2\pi}
\frac{1}{2} \sum_{\sigma=\pm 1} 
G_{\tilde \omega_n,\sigma}(p_y,p_z,x,x')
G_{-\tilde \omega_n,\sigma}(-p_y,-p_z,x,x')\nonumber \\
&\times&[\Delta^{t}•(\phi, p_z,x')+\Delta^t_{\tilde\omega_n}(x')-
i\sigma 
\tilde\Delta^t_{\tilde\omega_n}(x_1)],
\label{e20}
\end{eqnarray}
\begin{eqnarray}
&\tilde\Delta&^t_{\tilde\omega_n}(x)=nu^2\int dx' 
\int \frac{ dp_y}{2\pi} 
\int \limits_{-\pi/d}^{\pi/d}\frac{dp_z}{2\pi} 
\frac{1}{2} \sum_{\sigma=\pm 1} 
 G_{\tilde \omega_n,\sigma}(p_y,p_z,x,x')
G_{-\tilde \omega_n,\sigma}(-p_y,-p_z,x,x')\nonumber \\
&\times&(i\sigma(\Delta^{t}•(\phi, p_z,x')+\Delta^t_{\tilde\omega_n}(x'))
+
\tilde\Delta^t_{\tilde\omega_n}(x')].
\label{e21}
\end{eqnarray}

Taking into account that in common the paramagnetic limit of 
superconductivity $H_p$
is much higher than an orbital upper critical field we shall rest the 
general problem of influence of paramagnetism on superconductivity 
for a future investigations. In the absence of the Pauli paramagnetic 
interaction the equations (\ref{e16})-(\ref{e21}) are greatly 
simplified and we have the system of two equations
with equivalent structure for singlet 
and triplet pairing states
\begin{eqnarray}
&\Delta^{s,t}•(\phi, p_z,x)&=g \int dx'
\int \frac{dp'_y}{2\pi} 
\int \limits_{-\pi/d}^{\pi/d}\frac{dp'_z}{2\pi} 
\psi(\phi, p_z)\psi^*(\phi', p'_z)\nonumber \\
&\times& T\sum_n 
G_{\tilde \omega_n}(p'_y,p'_z,x,x')
G_{-\tilde \omega_n}(-p'_y,-p'_z,x,x')
(\Delta^{s,t}•(\phi', p'_z,x')+\Delta^{s,t}_{\tilde\omega_n}(x')),
\label{e22}
\end{eqnarray}
\begin{eqnarray}
&\Delta^{s,t}_{\tilde\omega_n}(x)&
=nu^2\int dx' 
\int \frac{dp_y}{2\pi} 
\int \limits_{-\pi/d}^{\pi/d}\frac{dp_z}{2\pi}\nonumber \\ 
&\times&G_{\tilde \omega_n}(p_y,p_z,x,x')
G_{-\tilde \omega_n}(-p_y,-p_z,x,x')
(\Delta^{s,t}•(\phi,\hat p_z,x')+\Delta^{s,t}_{\tilde\omega_n}(x')).
\label{e23}
\end{eqnarray}
Below we shall omit the supercripts $s,t$ using the common notation 
$\Delta^{s,t}(\phi,p_{z}•,x)=\Delta(\phi,p_{z}•,x)$, \\
$\Delta^{s,t}_{\tilde\omega_n}(x)=\Delta_{\tilde\omega_n}(x)$ for
the order parameters and the
self energy parts
both in singlet and in triplet case.

On this stage it is useful to note that the normal metal electron 
Green
function which we should find as a solution of the equation (\ref{e5})
depends of $p_y$ only through its square. Hence  for some 
unconventional 
superconducting phases like 
$\Delta(\phi,\hat p_z,x)= \sqrt 2(\sin^2\phi-\cos^2\phi)\Delta(x)$ 
for singlet pairing 
or $\Delta(\phi,\hat p_z,x)= \sqrt 2\cos \phi\Delta(x)$ for triplet 
pairing, the following property takes place
\begin{equation} 
\int \frac{dp_y}{2\pi}  
G_{\tilde \omega_n}(p_y,p_z,x,x')
G_{-\tilde \omega_n}(-p_y,-p_z,x,x')
(\Delta(\phi,\hat p_z,x)=0.
\label{e24}
\end{equation}
We will discuss further the only unconventional superconducting 
states obeying the
property (\ref{e24}).
For such the states the self energy part is equal to zero and we deal 
only with the order parameter equation.
\begin{eqnarray}
&\Delta(\phi, p_z,x)&=g \int dx'
\int \frac{dp'_y}{2\pi} 
\int \limits_{-\pi/d}^{\pi/d}\frac{dp'_z}{2\pi} 
\psi(\phi, p_z)\psi^*(\phi', p'_z)\nonumber \\
&\times& T\sum_n 
G_{\tilde \omega_n}(p'_y,p'_z,x,x')
G_{-\tilde \omega_n}(-p'_y,-p'_z,x,x')
\Delta(\phi', p'_z,x').
\label{e25}
\end{eqnarray}
The equality (\ref{e24}) is not valid for conventional 
superconducting state as well for many unconventional superconducting 
states where, as the consequence, there are nonzero selfenergy parts. 
In isotropic conventional superconducting state it  prevents a 
supression of the superconductivity by the ordinary impurities. For 
unconventional superconducting
states the presence of the self energy leads just to the mathematical 
complifications 
and does not change qualitatively the main results. 
  
The only difference of (\ref{e25}) from the pure case consists of 
change
$\omega_n \to \tilde \omega_n$ and one can use the expresion for the 
normal metal
electron Green
function found in the paper \cite{1}. 
It has nonzero value in the regions determined
by the inequality $\omega_n(x-x_1)\ge 0$ for positive value of $x$ 
component of electron momentum \\
($\alpha=1$) and by the inequality 
$\omega_n(x-x_1)\le 0$  for negative values of $x$ component momentum 
( $\alpha=-1$), where it is defined as 
\begin{eqnarray}
&G_{\tilde\omega_n}&(\phi,p_z,x,x_1)=\frac{-i{\it 
sign}\omega_n}{v_F\sin\phi}
\exp\left [-\frac{\alpha\tilde\omega_n(x-x_1)}{v_F\sin \phi}\right ]
\exp[i\alpha p_F(x-x_1)\sin \phi] \nonumber \\
\times &\exp&\left \{ \frac{i\alpha \lambda}{\sin \phi}
\sin\left [\frac{\omega_c(x-x_1)}{2v_F}\right ]
\cos\left [p_zd-\frac{\omega_c(x+x_1)}{2v_F}\right ]\right \}.
\label{e26}
\end{eqnarray}
Here $\lambda=4t/\omega_c$. 

This expression is valid under the condition 
\begin{equation}
|\sin\phi|>\sqrt{\frac{t}{\epsilon_F}}.
\label{e27}
\end{equation}
The disregard of the small intervals of the angle $\phi$
where $|\sin\phi|<(t/\epsilon_F)^{1/2}$ means that  
the only open trajectories of quasiparticles on the Fermi surface  
are taken into account \footnote{ The numerical calculation taking 
into account both
open and slosed trajectories, performed for clean case in the 
paper\cite{18}, just confirms the 
qualitative results of the article \cite{1} have been obtained in 
neglect of closed trajectories.} They give 
the main singular part to the kernels of the integral equation 
(\ref{e25}).

The substitution of (\ref{e26}) to the equation (\ref{e27}) gives 
after summation over the Matsubara frequency
\begin{eqnarray}
&\Delta(x)&=\tilde g\int\limits_{|x-x_1|>a\ sin \phi}dx_1
\int\limits_0^\pi\frac{d\phi}{2\pi v_F\sin \phi}
\int\limits_{-\pi}^{\pi}\frac{d(p_zd)}{2\pi}
\psi^*(\phi,p_z)\psi(\phi,p_z)
\frac{2\pi T\exp\left [-\frac{|x-x_1|}{l\sin \phi}\right ]}
{\sinh \left [\frac{ 2\pi T|x-x_1|}{v_F \sin \phi}\right ]}
 \nonumber \\
&\times &\exp\left \{\frac{2i\lambda}{\sin \phi}
\sin \left [\frac{\omega_c(x-x_1)}{2v_F}\right ]\sin(p_zd)
\sin \left [\frac{\omega_c(x+x_1)}{2v_F}\right ]\right \}
\Delta(x_1).
\label{e28}
\end{eqnarray}
Here $\tilde g=gm/4\pi d$ is dimensionless constant of pairing 
interaction,
$a$ is the small distances cutoff.
The equation (\ref{e28}) is the basic equation of the paper. The 
properties of its
solution we shall discuss in the next Section.

\section{The upper critical field}

Let us make an unessential simplification and consider $p_z$ 
independent superconducting states determined by functions 
$\psi(\phi)$ normalized as follows
$$
\int\limits_0^\pi\frac{d\phi}{\pi }
\psi^*(\phi)\psi(\phi)=1.
$$
In this case one can perform the integration over $p_z$
in (\ref{e28}):
\begin{eqnarray}
&\Delta(x)&=\tilde g\int\limits_{|x-x_1|>a\ sin \phi}dx_1
\int\limits_0^\pi\frac{d\phi}{2\pi v_F\sin \phi}
\psi^*(\phi)\psi(\phi)
\frac{2\pi T\exp \left [-\frac{|x-x_1|}{l\sin \phi}\right ]}
{\sinh \left [\frac{ 2\pi T|x-x_1|}{v_F \sin \phi}\right ]}
 \nonumber \\
&\times &{\cal I}_0 \left \{\frac{2\lambda}{\sin \phi}
\sin \left [\frac{\omega_c(x-x_1)}{2v_F}\right ]
\sin \left [\frac{\omega_c(x+x_1)}{2v_F}\right ]\right \}
\Delta(x_1).
\label{e29}
\end{eqnarray}
Here ${\cal I}_0(...)$ is the Bessel function. For the further 
purposes it is not important to work with the equation (\ref{e28}) or 
with (\ref{e29}) and for the determiness we will operate with the 
latter.

The equation (\ref{e29}) in the absence of a magnetic field
\begin{equation}
1=\tilde g\int\limits_{\frac{2\pi a T}{v_F}}^{\infty} 
\frac{dz}{\sinh z}\exp \left (-\frac {z}{2\pi T \tau}\right)
\label{e30}
\end{equation}
determines of the critical temperature $T_c$, which is expressed from 
here through 
the critical temperature $T_{c0}$ in a perfect crystal without 
impurities $l=\infty$
\begin{equation}
T_{c0}=\frac{v_F}{\pi a}\exp\left ( -\frac{1}{\tilde g}\right )
\label{e31}
\end{equation}
by means of well known relation
\begin{equation}
\ln \frac{T_{c}}{T_{c0}}=
\psi \left(\frac{1}{2}\right )-\psi 
\left(\frac{1}{2}+\frac{1}{4\pi\tau T_c}\right ).
\label{e32}
\end{equation} 
For the critical temperatures $T_c \sim T_{c0}$ the suppresion of the 
superconductivity by the impurities is: 
\begin{equation}
T_c=T_{c0}- \frac{\pi}{8\tau}.
\label{e33}
\end{equation}
One can also point out the condition of complete
suppresion of the superconductivity
\begin{equation}
\tau=\tau_c=\frac{\gamma }{\pi T_{c0}},
\label{e34}
\end{equation}
where $\ln \gamma =C=0,577...$ is the Euler constant.

The behavior of the upper critical field is determined by 
the relationship of the three spacial scales: $v_F/2\pi T$, $~l$ and 
$v_F/\omega_c$.  For the temperatures near to the zero field critical 
temperature  $T\approx T_c(H=0)$ the upper critical field 
$H_{c2}(T)$, or $\omega_{c2}(T)=eH_{c2}(T)v_Fd/c$ tends to zero and 
the inequality
\begin{equation}
min\left \{\frac{v_F}{2\pi T}, ~l\right 
\}~<~\frac{v_F}{\omega_{c2}(T)}
\label{e35}
\end{equation}
always presents \footnote {Near $T_c$ there is also formal solution 
of (\ref{e29}) with $\omega_{c2}>t$ ( see \cite{1}). This solution, 
however, is related to the region of magnetic fields,  out the 
region of validity of
the present theory  (\ref{e15}).} .
 
At the temperature decrease the upper critical field increase, but so 
long the inequality (\ref{e35}) takes place the essential interval of 
integration over $(x-x_1)$ in  (\ref{e29}) determined by the
$min(v_F/2\pi T, ~l)$. Hence,  one can use the smallness of 
$(x-x_1)v_F/\omega_c$
in the kernel of the equation  (\ref{e29}):
\begin{equation}
\Delta(x)=\tilde g\int\limits_{|x-x_1|>a\ sin \phi}dx_1
\int\limits_0^\pi\frac{d\phi}{2\pi v_F\sin \phi}
\psi^*(\phi)\psi(\phi)
\frac{2\pi T\exp \left [-\frac{|x-x_1|}{l\sin \phi}\right ]}
{\sinh \left [\frac{ 2\pi T|x-x_1|}{v_F \sin \phi}\right ]}
{\cal I}_0 \left \{\frac{4t(x-x_1)}{v_F\sin \phi}
\sin \frac{\omega_c x}{v_F}\right \}
\Delta(x_1).
\label{e36}
\end{equation}
The solution of this equation gives the correct value of 
$\omega_{c2}(T)$ so long the  inequality (\ref{e35}) is truth. By the 
substitution of the  $\omega_{c2}(T)$ 
obtained from this equation to the inequality (\ref{e35}) one can 
establish
the upper limit of the superconductor purity at which the   
$H_{c2}(T)$
found from this equation represent the correct value of the upper 
critical field 
up to zero temterature.

For the pure enough samples and low enough temperatures it can be 
turn out that
the opposite to the (\ref{e35}) relationship 
\begin{equation}
\frac{v_F}{\omega_{c2}(T)}~<~min\left \{\frac{v_F}{2\pi T}, ~l\right 
\}
\label{e37}
\end{equation}
breaking the correctness of transfering from (\ref{e29}) to 
(\ref{e36}) is realized.
It should be noted that ultraclean case demands a special 
investigation (see condition (\ref{e15}) and discussion in the paper
\cite {17}). One can claim however, that
at the temperatures
$$\frac{v_{F}•}{2\pi l}~<~T~<~\frac{\omega_c(T)}{2\pi}$$
the magnetic field dependence of the critical 
temperature in the equation (\ref{e29}) starts disappear or, in other 
words, the tendency to the divergency of the upper critical field 
pointed
out in the paper \cite{1} appears.

To solve of the equation (\ref{e36}) at arbitrary temperature and 
purity is possible only numerically. Here we shall discuss the case 
when  it allows the analitical solution. If the legth scale $\xi$, on 
which the function $\Delta(x)$ is lokalized, is larger than the 
essential distance of integration over $(x-x_1)$
\begin{equation}
\xi~>~min\left \{\frac{v_F}{2\pi T}, ~l\right \},
\label{e38}
\end{equation}
then one can expand 
$\Delta(x_1) \approx \Delta(x) +\Delta'(x)(x-x_1) 
+\Delta''(x)(x-x_1)^2/2$ under the integral in (\ref{e36}). Taking 
into consideration that under this condition
the argument of Bessel function turns to be small even on the upper 
boundary
of effective interval of the integration over $(x-x_1)$: 
\begin{equation}
\frac{t\omega_{c2}(T)\xi ~min\left \{\frac{v_F}{2\pi T}, ~l\right 
\}}{v_F^2}
\approx \frac{t ~min\left \{\frac{v_F}{2\pi T}, ~l\right 
\}}{\epsilon_F~\xi}
~<~1,
\label{e39}
\end{equation}
one can also expand Bessel function ${\cal I}_0(x)\approx 1-x^2/4$. 
As the result we get
the differential equation 
\begin{equation}
\left (\ln \frac{T_{c0}}{T}-\psi \left(\frac{1}{2}+\frac{1}{4\pi\tau 
T}\right )+
\psi \left(\frac{1}{2}\right )\right )\Delta (x)=
-\frac{C~I(\alpha)}{2}\left (\frac{v_F}{2\pi T}\right )^2 
\Delta''(x)~+~
I(\alpha)\left (\frac{t\omega_{c2}(T)x}{\pi v_F T}\right )^2 
\Delta(x),
\label{e40}
\end{equation} 
where $\alpha=(2\pi T\tau)^{-1}$,
\begin{equation}
I(\alpha)=\int \limits _{0}^{\infty} \frac{z^2dz}{\sinh 
z}\exp(-\alpha z)=
4\sum_{n=0}^{\infty}\frac{1}{(2n+1+\alpha)^3},
\label{e41}
\end{equation}
\begin{equation}
C_{\psi}=\int\limits _{0}^{\pi}\frac{d\phi}{\pi}\psi^*(\phi) 
\psi(\phi)\sin^2\phi.
\label{e42}
\end{equation} 

In pure case $\alpha\approx\alpha_c= (2\pi T_{c}\tau)^{-1}\ll 1$
the inequality (\ref{e38}) is valid only in vicinity of the critical 
temperature.
Puting $T=T_c$ in the right hand side and taking its lowest eigen 
value we obtain
\begin{equation}
\ln \frac{T_{c0}}{T}-\psi \left(\frac{1}{2}+\frac{1}{4\pi\tau 
T}\right )+
\psi \left(\frac{1}{2}\right )=
\frac{\sqrt {C_{\psi}} I(\alpha_c)t\omega_{c2}(T)}{2\sqrt 2 \pi^2 
T_c^2}.
\label{e43}
\end{equation}
Summing this equation with equation (\ref{e32}) and leaving only 
linear on $T-T_c$
and on the impurity concentration terms we get 
\begin{equation}
\omega_{c2}(T)=\frac{ev_Fd}{c}H_{c2}(T)=
\frac{4\sqrt 2 \pi^2 }{7\zeta (3)\sqrt {C_{\psi}} t}
\left (T_{c0}-\beta\frac{\pi }{8\tau}\right )(T_{c} -T).
\label{e44}
\end{equation}
Here the coefficient
\begin{equation}
\beta=2-\frac{90\zeta(4)}{7\pi^2\zeta(3)}\approx 0.83,
\label{e45}
\end{equation}
shows that, the slop of $H_{ c2}(T)$ at $T=T_c$
decreases with the increase of the impurity concentration somewhat 
slower than
$T_c$ itself (see eqn. (\ref{e33})) .

The equation  (\ref{e36})  does not contain any divergency of 
$H_{c2}(T)$
and the expression (\ref{e44}) which is valid in Ginzburg-Landau 
region
can be used at arbitrary temperature as the  estimate of upper 
critical field
from above. Hence, to establish the limits of a sample purity,
at which the equation  (\ref{e36}) works, one may substitute 
(\ref{e44}) at zero
temperature into the inequality (\ref{e35}). Omitting the numerical 
factor of the order of unity we have 
\begin{equation}
l~<~\frac{t}{T_c}\xi_{ab},
\label{e46}
\end{equation}
where $\xi_{ab}=v_F/2\pi T_c$ is the basal plane coherence length. We 
see, that there is the good reserve in sample purity
in the limits of which one can not expect low temperature divergency 
of the upper critical field.  In the $Sr_2RuO_4$ the mean free path 
should be approximately 10
times larger than the basal plane coherence length $\xi_0$ to go out 
limit (\ref{e46}).

Let us consider now the dirty case: $T_c\ll T_{c0}$, $\alpha\gg 1$ 
and 
$I(\alpha)\approx \alpha^{-2}$ allowing analytical solution for 
$H_{c2}(T)$
at arbitrary temperature. Taking the lowest eigen value of the 
equation
 (\ref{e40})
we get 
\begin{equation}
\ln \frac{T_{c0}}{T}-\psi \left(\frac{1}{2}+\frac{1}{4\pi\tau 
T}\right )+
\psi \left(\frac{1}{2}\right )=
\sqrt {2C_{\psi}} \tau^2 t \omega_{c2}(T).
\label{e47}
\end{equation}
Summing  of this equation with equation (\ref{e32}) yields 
\begin{equation}
\omega_{c2}(T)=\frac{\ln \frac{T_c}{T}-\psi 
\left(\frac{1}{2}+\frac{1}{4\pi\tau T}\right )+\psi 
\left(\frac{1}{2}+\frac{1}{4\pi\tau T_c}\right )}
{\sqrt {2C_{\psi}} \tau^2 t} .
\label{e48}
\end{equation}
This expression is correct at any temperature. One can rewrite it 
approximatively
in more simple form
\begin{equation}
\omega_{c2}(T)=\frac{\sqrt 2 \pi^2}{3\sqrt {C_{\psi}} t}(T_c^2-T^2) ,
\label{e49}
\end{equation}
where
\begin{equation}
 T_c=\frac{1}{\pi \tau}
\left ( \frac{3}{2}\ln \frac{\pi T_{c0}\tau}{\gamma}\right )^{1/2}.
\label{e50}
\end{equation}

\section{Conclusion}
The system of linear integral equations for the order parameter of 
conventional and unconventional superconducting state in a layered 
crystals 
with impurities under magnetic field parallel to the conducting 
layers is derived. It is shown that so long the purity of the samples 
does not exceed high enough level  (\ref{e46}) there is no tendency 
to the low temperature divergency of the upper critical field. The 
analytical solution of the equations in the clean
(Ginzburg-Landau region) and dirty (arbitrary temperature) limits are 
presented.

\section*{Acknowledgements}
I express my best gratitudes to Dr. Manfred Sigrist and all the 
members of
Condensed matter theory group at Yukawa Institute for Theoretical 
physics
for their kind hospitality and interest to my work during my stay in 
Kyoto
autumn 1999 where the significant part of this work have been 
completed.

I also indebted to Dr. Yoshiteru Maeno and his collaborators having 
stimulated
my interest to the problem of upper critical field in layered 
superconductors. 

I appreciate the valuable remarques of prof. M.Walker,
 prof. P.Nozieres and prof. Yu.A.Bychkov  have resulted 
in the improvement in the initial text of the article.
\newpage


\end{document}